\documentclass{nature}
\usepackage{amssymb}
\usepackage[utf8]{inputenc}
\usepackage[]{graphicx,bm,color,subfigure,amsmath}
\usepackage{graphicx}
\usepackage{caption}
\usepackage{gensymb}
\usepackage{upgreek}

\usepackage{savesym}
\savesymbol{spacing}

\usepackage{setspace}

\usepackage[export]{adjustbox}
\usepackage[utf8]{inputenc}
\usepackage[unicode=true]{hyperref}
\usepackage{hyperref}
\usepackage{upgreek}

\makeatletter
\let\saved@includegraphics\includegraphics
\AtBeginDocument{\let\includegraphics\saved@includegraphics}
\renewenvironment*{figure}{\@float{figure}}{\end@float}
\makeatother

\title{Coexisting and interacting spin torque driven free and reference layer magnetic droplet solitons\\
}

\author{S. Jiang$^{1,2,3\ast}$, S. Chung$^{2,4,\ast}$, M. Ahlberg$^{2,\ast}$, A. Frisk$^{2}$, Q. Tuan Le$^{2,5}$, H. Mazraati$^{2}$, A. Houshang$^{2}$, O. Heinonen$^{6,\dag}$ \& J. \AA{}kerman$^{2,3,5}$}

\begin{document}

\maketitle

\begin{affiliations}
 \item School of Microelectronics, South China University of Technology, 511442 Guangzhou, China
 \item Physics Department, University of Gothenburg, 412 96, Gothenburg, Sweden
 \item NanOsc AB, 164 40 Kista, Sweden
 \item Department of Physics Education, Korea National University of Education, Cheongju 28173, Korea
\item Department of Applied Physics, School of Engineering Sciences, KTH Royal Institute of Technology, 100 44 Stockholm, Sweden
 \item Materials Science Division, Argonne National Laboratory, Lemont, IL 60439, USA

$^\ast$These authors contributed equally to this work.
$^\dag$Present and permanent address: Seagate Technology, 7801 Computer Ave., Bloomington, MN 55435 
\end{affiliations}

\begin{abstract}
Magnetic droplets are nanoscale, non-topological, magnetodynamical solitons that can be nucleated in spin torque nano-oscillators (STNOs) or spin Hall nano-oscillators (SHNOs). All theoretical, numerical, and experimental droplet studies have so far focused on the free layer (FL), and any additional dynamics in the reference layer (RL) have been entirely ignored. Here we show, using all-perpendicular STNOs, that there is not only significant magnetodynamics in the RL, but the reference layer itself can host a droplet coexisting with the FL droplet. Both droplets are observed experimentally as stepwise changes and sharp peaks in the dc and differential resistance, respectively. Whereas the single FL droplet is highly stable, the coexistence state exhibits high-power broadband microwave noise. Micromagnetic simulations corroborate the experimental results and reveal a strong interaction between the droplets. Our demonstration of strongly interacting and closely spaced droplets offers a unique platform for fundamental studies of highly non-linear soliton pair dynamics.
\end{abstract}

Magnetization dynamics can be excited by spin-transfer-torque (STT) and/or spin-orbit-torque (SOT) in spintronic nanodevices\cite{Mohseni2013a,Chen2016,Tarequzzaman2019,Song2020}. Using nano-oscillators, rich dynamics can be generated, such as propagating spin waves\cite{Houshang2018}, localized bullets\cite{Bauer1998, dumas2013prl}, magnetic droplets\cite{Mohseni2013a}, dynamic vortices\cite{Pribiag2007} and skyrmions\cite{Garcia-Sanchez2016}. Their highly tunable dynamics are essential for applications in radio-frequency electronics\cite{Choi2014,Sharma2021}, magnetic random access memory (MRAM)\cite{Okamoto2015}, magnonics\cite{Barman_2021}, neuromorphic computing\cite{Romera2018,Zahedinejad2020}, and Ising machines\cite{Albertsson2021}. 

The magnetic droplet soliton has been the subject of numerous studies, covering theoretical\cite{slavin2009,Hoefer2010,Hoefer2012,Moore2019}, numerical\cite{Zhou2015,Xiao2016,Mohseni2018,Moore2019,Zheng2020}, and experimental aspects\cite{Mohseni2013a,Macia2014,Chung2016,Chung2018PRL,Shi2022,Ahlberg2022}. However, all droplet studies so far have neglected the magnetodynamics of the reference layer. On the other hand, problems with back-hopping in magnetic switching have been addressed for more than a decade.\cite{Hou2011} Back-hopping occurs when the applied dc current is substantially higher than the critical switching current, which results in the destabilization of the reference layer and excitation of dynamics.\cite{Min2009,Kim2016,Abert2018,Chen2020,Devolder2020} RL modes have also been observed in magnetic tunnel junctions.\cite{ Muduli2011} Therefore, both the free and the reference layers are expected to demonstrate dynamics when the applied current is high enough. 

\subsection{Experimental characteristics of double droplets.}

\begin{figure}[htb]
\includegraphics [width=1\textwidth]{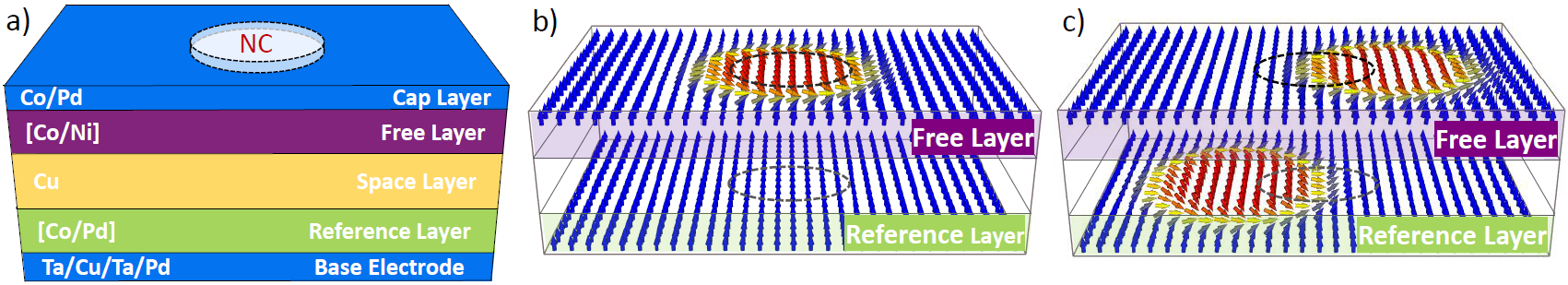}
\centering
    \caption{\textbf{Sample schematics.} \textbf{a} Schematic of an STNO device with stack information. Both the free (Co/Ni multilayer) and the reference layer (Co/Pd multilayer) have strong perpendicular magnetic anisotropy. \textbf{b} Single droplet nucleated in the free layer. \textbf{c} Double droplets in both the free and fixed layers. Dashed circles indicate the nanocontact (NC) areas.}
    \label{fig1}
\end{figure}

We use nanocontact STNOs, with strong perpendicular magnetic anisotropy (PMA) in both the 
free and reference layer. Figure~\ref{fig1} shows the layer materials and order in the stack (Fig.~\ref{fig1}a) along with illustrations of a single FL droplet (Fig.~\ref{fig1}b). While the RL is usually considered as simply a polarization layer, in essence without dynamics. we show that both the RL and FL can hold droplets, which coexist as illustrated in Fig.~\ref{fig1}c.

Droplet nucleation is often identified by a sharp drop in the detected frequency. However, in our measurement,
the applied field is directed along the PMA axis of both layers. As a consequence, the resistance of the STNO is unaffected by the in-plane magnetization direction, and it is therefore not possible to use GMR to experimentally harvest the microwave frequency precession of the spins in the droplet perimeter. Instead, we use the absolute (DC) and differential resistance ($R_{\text{DC}}$, $R_{\text{dV/dI}}$, respectively) together with the power spectral density (PSD) to identify the different phases. A similar approach was employed in an earlier study where we observed the transition from a droplet to a static bubble\cite{Ahlberg2022}.

\begin{figure}[htb]
  \includegraphics[width=1\textwidth]{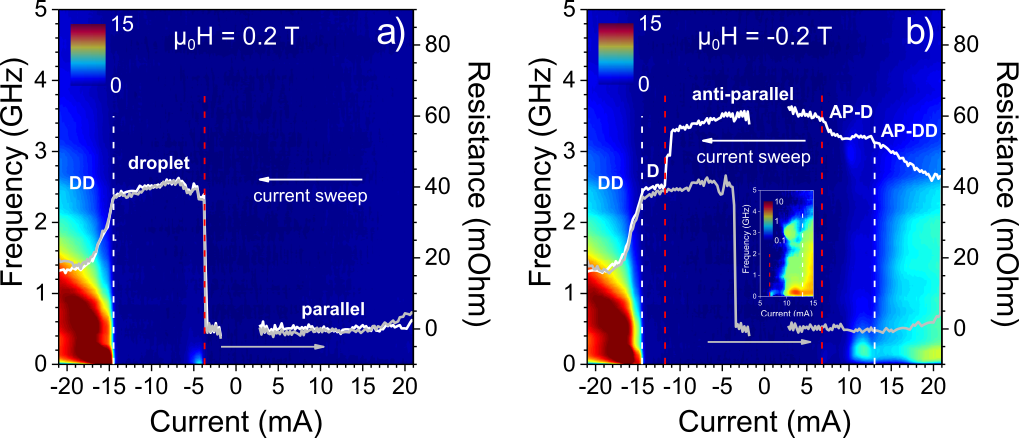}
\centering
    \caption{\textbf{DC and RF characterization.} The dc resistance (white line) and corresponding PSD (color map) of a current sweep from $I=21$~mA to $I=-21$~mA, at \textbf{a} $\upmu_{0} H=+0.2$~T and \textbf{b} $\upmu_{0} H=-0.2$~T. The gray lines show the resistance of the backward sweeps $I=-21 \rightarrow 21$~mA; the corresponding PSDs are not shown. AP, D, and DD denote Anti-Parallel, single Droplet, and Double Droplets, respectively. A constant device resistance and a parabolic background (caused by Joule heating) have been removed from the raw data, thus the resulting parallel resistance is zero. The inset in \textbf{b} shows the PSD at $I= 5$--$15$~mA on a logarithmic scale. }
    \label{fig2}
\end{figure}

Figure~\ref{fig2} demonstrates the experimental features of all magnetic states observed: simple static parallel (P) and anti-parallel (AP) alignments of the FL and RL, single FL droplet (D), single FL droplet in the AP state (AP-D), and the novel states of double droplets in the P (DD) and the AP (AP-DD) states. A positive applied current ($I$) has no effect on parallel layers (Fig.~\ref{fig2}a), since the STT adds to the damping-like torque in this case, giving a virtually constant resistance. When the current polarity is reversed, an ordinary droplet is nucleated at $I \approx -4$~mA, which is manifested by a clear step in $R_{\text{DC}}$ and a very weak microwave signal\cite{Chung2016, Chung2018PRL} in a narrow current region just after nucleation. The resistance remains high for a range of about 10~mA without any concomitant noise, which means the droplet is remarkably stable. 

Double droplets, i.e. the emergence of a coexisting droplet in the reference layer, appear as the current is further increased. The transition is identified by a dramatic onset of low-frequency noise, while the resistance decreases. The microwave noise is likely caused by droplet annihilation/re-nucleation and drift, indicating much less stability of the DD state. In addition, if both droplets were stable, the resistance should approach the parallel state value. Instead, it levels out at about $17$~m$\ohm$, approximately halfway between the P and D states.

We now turn to the opposite polarity of the field, $\upmu_{0}H=-0.2$~T (Fig.~\ref{fig2}b). At this field, the sample is in an AP state and droplet nucleation is facilitated by positive currents. Although the sweep starts at $I=+21$~mA, we start describing lower currents. The resistance shows a small drop at $I = 6.8$~mA, corresponding to droplet formation. The transition is also accompanied by measurable microwave noise, although it is hard to discern in the figure. The inset highlights the small signal by presenting the PSD on a logarithmic scale. The decrease in resistance ($\Delta R_{\text{D}}^{\text{AP}} \approx 6$~m$\ohm$) is lower than the corresponding increment of $R$ for a droplet in a parallel setting ($\Delta R_{\text{D}}^{\text{P}} \approx 40$~m$\ohm$). The difference in $\Delta R_{\text{D}}$ translates into a difference in size, and the smaller size of the AP-droplet is consistent with the effect from Zhang-Li torque (ZLT)\cite{Chung2018PRL}. The AP-droplet becomes unstable at higher currents of about $10$~mA, and ultimately a second droplet is nucleated in the reference layer (AP-DD) at $13$~mA as indicated by the falling resistance and the high noise level. 

The behavior at negative currents is identical to $\upmu_{0}H=+0.2$~T (Fig.~\ref{fig2}a) with one important exception: at a threshold current the magnetic state switches from AP to P. Since the RL magnetization is anti-parallel to both the FL\cite{Abert2018} and the applied field, this layer becomes increasingly unstable with increasing current and finally reverses. The reversal is observed by a marked decline in $R_{\text{DC}}$, and a stable ordinary droplet forms without any concomitant microwave signal. The single droplet phase is followed by RL-droplet nucleation (DD), and the backward sweep (gray line) repeats the same features. However, at the highest positive current (very last data point, $I=21$~mA) the STT forces an anti-parallel alignment, and AP-double-droplets once again appear. 

The results reveal large differences in the experimental traces of droplets in the parallel state compared to the AP state. In a first approximation, they should be the same. However, as already mentioned, ZLT contracts the AP-droplet, and the very presence of ZLT is caused by the current distribution having an in-plane component\cite{Chung2018PRL}. The symmetry of the system is thus broken by the applied current. This not only causes unequal droplet sizes, but also influences the evolution in the double droplet regime. The effect on the observed PSD is remarkable. The maximum power of the P-DD is roughly twice as high compared to the AP-DD, as seen by comparing the dynamic signal in Fig~\ref{fig2}b at negative and positive currents, respectively. The shape of the signal is also different. At negative currents, the power gradually levels off with frequency. At positive currents, the signal stays constant over a range of 2~GHz. For both conditions, the power diminishes as the frequency approaches 5~GHz. Consequently, the drift motion of coexisting droplets is highly dependent on the initial orientation of the magnetic layers. Nevertheless, the dominating time scale is similar. The PSDs of both the P and AP states show a maximum at about 0.2~GHz, and no signal above 5~GHz. 

\subsection{Current--field phase diagram.}

\begin{figure}[t]
  \includegraphics[width=1\textwidth]{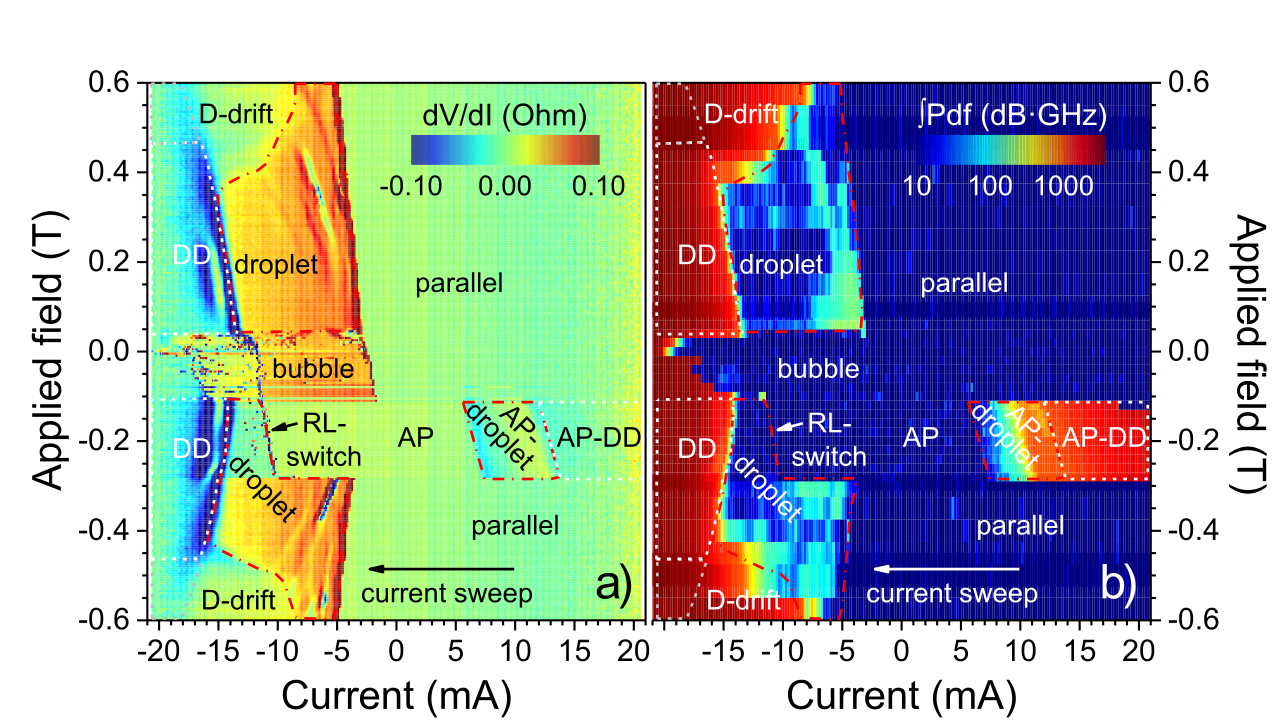}
\centering
    \caption{\textbf{Phase diagrams based on the STNO differential resistance and microwave noise.} \textbf{a} STNO differential resistance (dV/dI) as a function of applied current and field; a parabolic background has been removed. \textbf{b} Microwave noise, integrated over 0--5~GHz, presented on a logarithmic scale. The red dash-dotted lines mark regions of stable single droplets, while areas with double droplets (DD) are indicated by white dashed lines. The gray dashed lines denote sectors where it is hard to distinguish single droplet drift (D-drift) and double droplet dynamics. Droplets in the anti-parallel (AP) state are found at positive currents. Between each current sweep from $21$ to $-21$~mA (displayed) and back again (not shown), the 
sample was first saturated in a field of $0.6$~T, whereafter the measurement field was set.
 The data in \textbf{a} and \textbf{b} are taken from two different measurements.}
    \label{fig3}
\end{figure}

The phase diagram displayed in Fig.~\ref{fig3}a is constructed from $R_{\text{dV/dI}}$ and Fig.~\ref{fig3}b presents microwave noise, integrated over 0--5~GHz, using a logarithmic scale to highlight small signals. Different magnetic states are easily identified by comparing the two subfigures.

The only visible features in positive currents are found in the anti-parallel configuration ($-0.28$~T $<$$\upmu_{0}H$$<$$-0.10$~T), where double droplets nucleate immediately as a large $+I$ is applied. As the current is reduced, the RL droplet first vanishes, followed by the annihilation of the single FL droplet, leaving a simple AP state. In this state, negative currents destabilize the reference layer\cite{ Abert2018, Choi2016} and at around $I=-10$~mA the RL switches into the direction of the applied field. The differential resistance does not reveal the subsequential nucleation of a droplet, and there is no associated microwave noise. However, the magnitude of $R_{\text{DC}}$ after the switch is identical to an ordinary droplet (Fig.~\ref{fig2}a) and we conclude that the free layer indeed hosts a droplet once the RL is reversed. Further increase of the current nucleates a droplet in the reference layer as well, and this transition is clearly visible in Fig.~\ref{fig3}a as negative peaks (blue lines), and in Fig.~\ref{fig3}b as a distinct onset of strong noise.

The single droplet nucleation boundaries follow the linear dependence expected by STT-theory\cite{Slonczewski1996}. Figure~\ref{fig3} shows that the droplet is exceedingly stable in a wide range of fields and currents, which we have indicated by red dash-dotted lines. The small microwave noise signal observed in this range is related to droplet mode conversions also visible as peaks in $R_{\text{dV/dI}}$. A more well-defined map of the noise at low currents is found in Ref.~[\citen{Ahlberg2022}]. The range where double droplets are unambiguously present is marked by white dashed lines in Fig.~\ref{fig3}. The transition is more blurred above $\upmu_{0} \left| H \right| \approx 0.46$~T, and the FL droplet is also accompanied by strong microwave noise at high fields and currents. The range where it is difficult to distinguish single and double droplets is indicated by gray lines.

\subsection{Microwave noise characteristics.}

\begin{figure}[t]
    \includegraphics[width=1\textwidth]{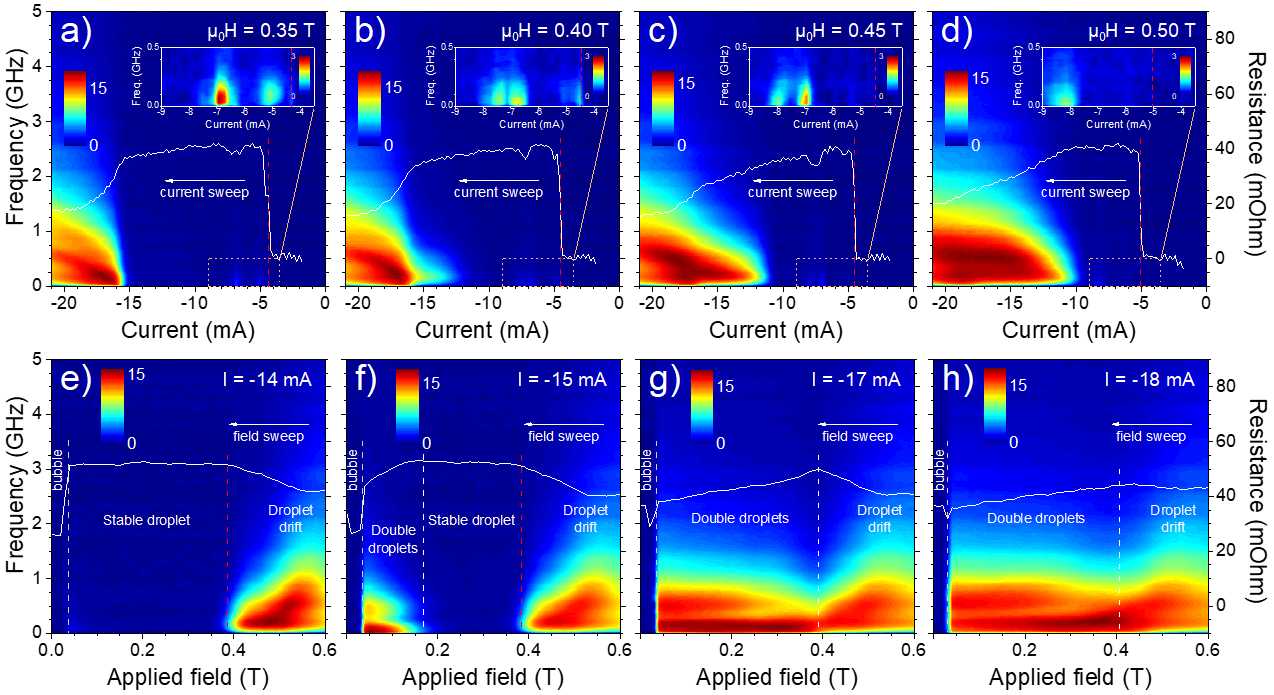}
\centering
    \caption{\textbf{Microwave noise PSD and DC resistance at selected fields and currents.} \textbf{a--d} PSD (color map) and dc resistance (white lines) of current sweeps at four different fields: \textbf{a} $\upmu_{0} H=0.35$~T, \textbf{b} $\upmu_{0} H=0.40$~T, \textbf{c} $\upmu_{0} H=0.45$~T, and \textbf{d} $\upmu_{0} H=0.50$~T. The insets zoom in on small signals in the PSD; the displayed area is given by the orange dashed box. The red dashed lines mark the nucleation of a single droplet. The current was swept from $I=+21$~mA to $I=-21$~mA, but only data for negative currents is shown, since nothing noteworthy occurs at positive $I$.  
\textbf{e--h} PSD (color map) and dc resistance (white lines) of field scans at four different currents: \textbf{e} $I=-14$~mA,  \textbf{f} $I=-15$~mA, \textbf{g} $I=-17$~mA, \textbf{h} $I=-18$~mA. A parabolic background is removed from the resistance.}
    \label{fig4}
\end{figure}

Figure~\ref{fig4} shows the dc~resistance and microwave noise at four selected fields and currents. The behavior at $\upmu_{0}H=0.35$~T displayed in Fig.~\ref{fig4}a is very similar to the observation at $\upmu_{0}H=0.20$~T (Fig.~\ref{fig2}a). The distinct step in resistance is a sign of the nucleation of a single droplet, which remains stable for a wide current range until double droplets appear at $I \approx -15$~mA as revealed by the decline of $R_{\text{DC}}$ together with the onset of strong noise. In Fig.~\ref{fig4}b the field is increased by $0.05$~T and the image changes slightly. The microwave noise appears at a weaker current, and covers initially lower frequencies compared to the double droplet signal. The overall resistance profile remains virtually unchanged. It is still easy to find a distinction between the single and DD phases, and we attribute the initial noise to single droplet drift. As the field is further increased in Fig.~\ref{fig4}c-d, the resistance acquires a noticeable slope, and the introduction of noise moves to even weaker $I$. The distinction between single and double droplets is smeared out.

The field sweeps presented in Fig.~\ref{fig4}e-h give a more explicit illustration of the single and double droplets' characteristics and the gradual fading of the border between them. At moderate currents, represented by $I=-14$~mA in Fig.~\ref{fig4}e, there are only free layer droplets, which freeze into static bubbles close to zero field\cite{Ahlberg2022}. The droplet experiences drift at high fields, manifested by a reduction of the time-averaged resistance and the presence of microwave noise. A comparable drift is observed at the same fields for $I=-15$~mA, but Fig.~\ref{fig4}f also unveil the difference between drift noise compared to double droplet noise. The dynamic signal of double droplets covers a lower frequency range and has a two-peak-like shape, while single droplet drift noise diminishes close to zero GHz. Furthermore, $R_{\text{DC}}$ has a positive slope in the DD regime, stays constant for a stable ordinary droplet, and descends in presence of drift. Figure~\ref{fig4}g shows that the different features are still visible at $I=-17$~mA, although there is no clear cut between one and another. For even higher currents, exemplified in Fig.~\ref{fig4}h, there is no well-defined aspect that defines the two regimes, although the close-to-zero frequency intensity decline at higher fields, indicating single droplet drift.

The insets in Fig.~\ref{fig4}a-d zoom in onto selected currents and emphasize low noise signals in the single droplet phase. The nucleation of a droplet is not always accompanied by measurable dynamics. However, transitions between (single) droplet modes are discernable by noticeable kinks in $R_{\text{DC}}$ together with evident low-frequency signals. These mode transitions are also found in Fig.~\ref{fig3} and the associated noise has no intensity above $\approx 0.5$~GHz. The lateral dynamics of mode transitions is consequently much slower compared to the movements related to single droplet drift or double droplet interactions.

\subsection{Micromagnetic simulations.}

We have performed micromagnetic simulations to further explore the double droplet phase. Magnetodynamics in both the free and reference layers were simulated simultaneously. Taking the back-hopping effect into account, we considered the real-time magnetization of one layer to serve as the polarization layer of the other layer. Zero-temperature, i.e. $T=0$~K, was used in all simulations. Overall, the outcome confirms the experimental observation of droplets in the reference layer. Furthermore, the double droplet state can, based on its dynamic signature, be divided into three subcategories---periodic, pseudo-periodic, and chaotic. 

\begin{figure}
\includegraphics [width=1\textwidth]{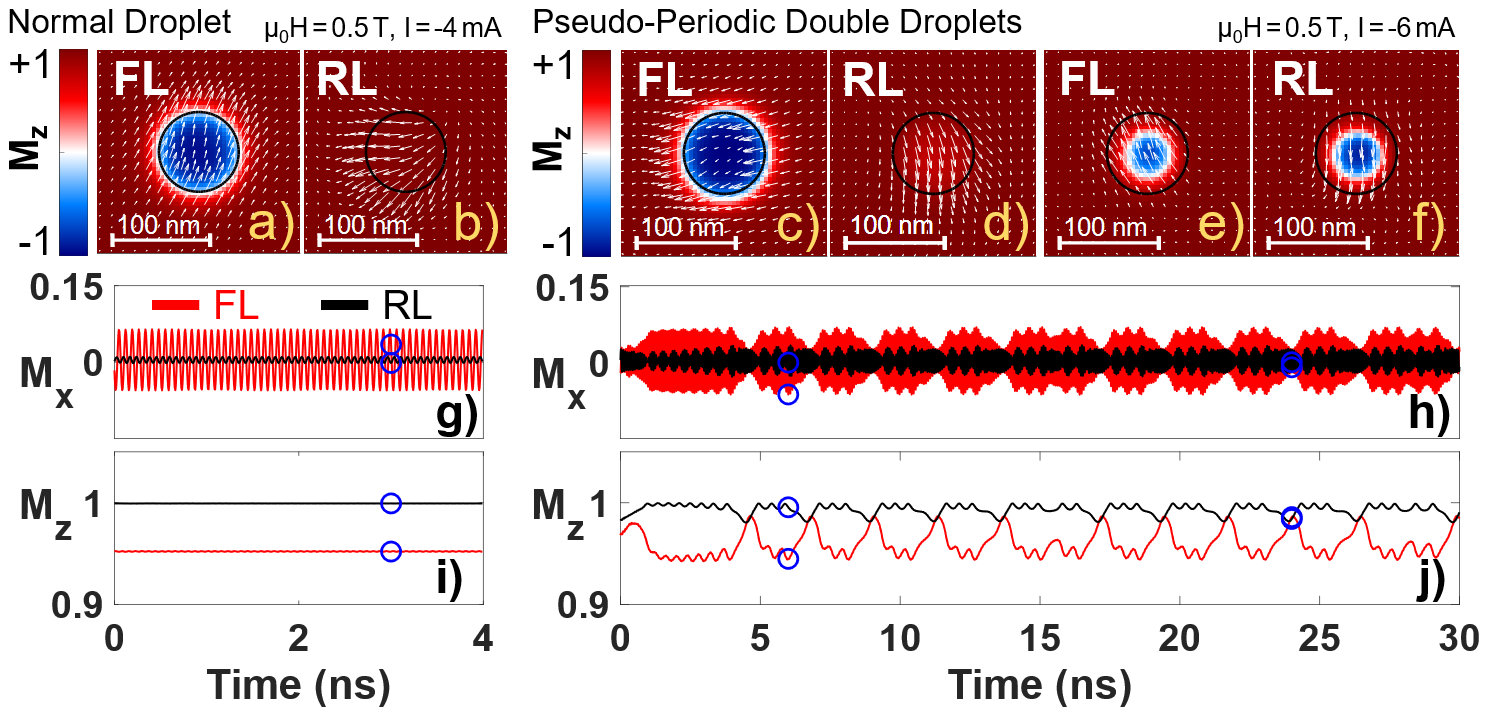}
\centering
    \caption{\textbf{Ordinary free layer droplet and pseudo-periodic double droplets.} \textbf{a--f} Droplet snapshots together with \textbf{g--j} the time evolution of $M_{\text{x}}$ (in-plane) and $M_{\text{z}}$ (out-of-plane). The magnetization of the free layer (FL) and the reference layer (RL) are given by red and black lines, respectively. The blue circles in \textbf{g--j} show the corresponding time and magnetization of the snapshots.}
    \label{fig5}
\end{figure}

An ordinary droplet is formed at high field and low current ($\upmu_{0} H = 0.5$~T, $I=-4$~mA) as illustrated in Fig.~\ref{fig5}a-b, which show snapshots of the magnetic states in the free and reference layers, respectively. In addition, the $M_{\text{x}}$- and $M_{\text{z}}$-components of the two layers are displayed as a function of time in Fig.~\ref{fig5}g-i. The OOP magnetization is basically constant with time and no drift is observed. The in-plane magnetization manifests tiny effects in the RL with a frequency equal to the uniform FL droplet precession, which is 14~GHz. This means the droplet precesses at the theoretical lower bound, i.e. the Zeeman frequency ($f_{\text{Z}} = \gamma/2 \pi \upmu_{0} H$). However, the frequency of a dissipative droplet is predicted to be a monotonically decreasing function of its diameter and to only reach the lower bound in the limit of infinite size\cite{Hoefer2010}. Our droplet is not particularly large, and the low frequency must be attributed to interactions with the reference layer, which slow down the precession.

Double droplets emerge as the field is kept constant and the current is increased. Figure~\ref{fig5}c-j presents the characteristics of this state. Neither the FL droplet, nor the RL droplet is stable with time. The evolution of the droplet volume is mirrored by the $M_{\text{z}}$-component, see Fig~\ref{fig5}j. Starting at $t = 1$~ns, the free layer droplet is fully developed, while the RL only demonstrates small wiggles. Similar wiggles are seen in the free layer. Figure~\ref{fig5}c-d illustrates the presence of an FL droplet accompanied by a small RL perturbation. The reference layer excitation grows larger with time and after a certain interval the RL droplet starts to grow ($t \approx 4$~ns). Simultaneously, the FL droplet contracts. As the droplets reach about the same size (Fig.~\ref{fig5}e-f), the FL droplet begins to expand and the RL droplet diminishes. Once the RL droplet vanishes, new wiggles appear, and the process starts over.   

The pseudo-periodic state DD is thus characterized by an expansion/contraction process. It looks periodic at first glance, but the periodicity is far from perfect. Occasionally, the process is disrupted before the droplets reach their minimum/maximum size, see e.g. Fig.~\ref{fig5}j at $t = 15$~ns. A stochastic element is clearly involved in the process. Moreover, the intervals between subsequential peaks in $M_{\text{z}}$ are not identical. The fast precessions of the $x$-components also exhibit both beating and a gradual phase shift. The stochastics must be driven by intrinsic non-linear dynamics governed by the underlying equations, and/or numerical noise, since $T=0$~K rules out any thermal effects. Albeit unstable, the dynamics comprises different timescales that can be estimated. The dominating in-plane precession frequency is similar to the single droplet state, $\approx 14$~GHz. The small wiggles in both components occur with a period of $\approx 0.45$~ns. The large peaks in $M_{\text{z}}$ emerge roughly every two nanoseconds, while the process is disrupted once in $10$--$15$~ns.

\begin{figure}
  \includegraphics[width=1\textwidth]{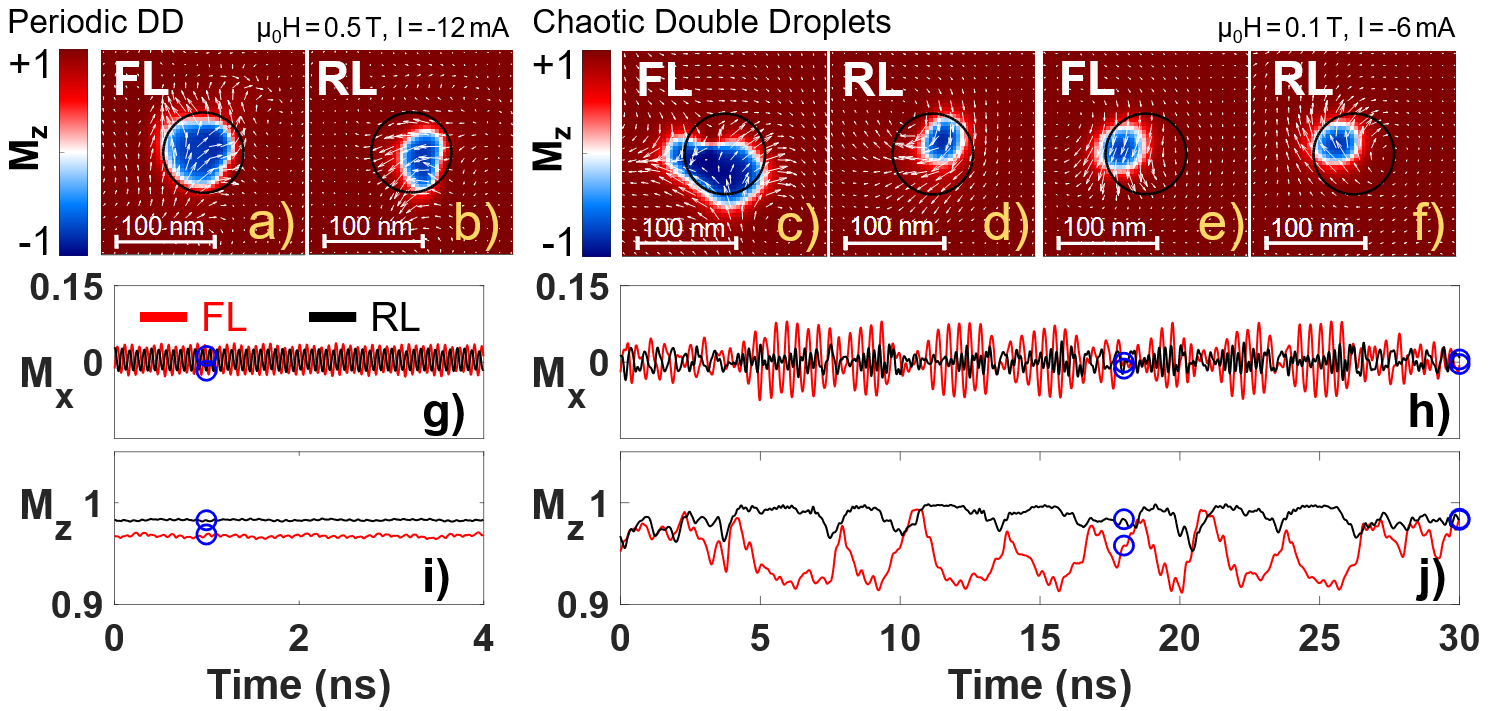}
\centering
    \caption{\textbf{Periodic and chaotic double droplets.} \textbf{a--f} Droplet snapshots together with \textbf{g--j} the time evolution of $M_{\text{x}}$ (in-plane) and $M_{\text{z}}$ (out-of-plane). The magnetization of the free layer (FL) and the reference layer (RL) are given by red and black lines, respectively. The blue circles in \textbf{g--j} show the corresponding time and magnetization of the snapshots.}
    \label{fig6}
\end{figure}

A periodic DD state develops when the current is further increased, see Fig.~\ref{fig6}. Here, the free layer droplet is always accompanied by a reference layer counterpart. Each droplet gyrates with a frequency of $\approx 2.4$~GHz, around a point close to the NC center, and they expel each other. Their interdependence leads not only to a noticeable slowing down, $f=12.7$~GHz, but also that both precess with nearly equal frequency. The sub-Zeeman frequency can be rationalized keeping the magnetic structure in mind. The periodic double droplets exhibit very divergent in-plane textures (Fig.~\ref{fig6}a,b), while a homogenous magnetic state is necessary to interpretate $f_{\text{Z}}$ as the lowest accessible frequency. The mutual precession frequency is on the other hand unexpected from single droplet theory. The layers have different anisotropy and saturation magnetization, but more importantly radically different degrees of spin reversal (mirrored by $M_{\text{z}}$ in Fig.~\ref{fig6}i). The fact that the unequal properties do not affect the frequency, highlights the importance of the second droplet. 

A chaotic state appears upon lowering both field and current ($\upmu_{0} H = 0.1$~T, $I=-6$~mA), as clearly seen in Fig.~\ref{fig6}h-j. Both droplets are unstable and for short moments the FL-D practically vanishes, while the RL-D is missing for extended periods of time. There is no correlation between the $M_{\text{z}}$-components of the two droplets, in contrast to the pseudo-periodic case. Besides, the in-plane precession is uniform only in the absence of an RL droplet. 

The micromagnetic results compare well with the experiments, given the simple model. The stable FL-droplet phase occur at low currents, while both the pseudo-periodic and chaotic double droplets display substantial oscillations in $M_{\text{z}}$, which corresponds to an experimentally measurable signal. The oscillations appear on time scales on the order of several to tenths of nanoseconds, where the largest amplitudes are found for the longest cycles. These time scales match the experimental microwave maxima around 0.15~GHz, as well as the broad falloff towards higher frequencies. 

The periodic double droplets are not found in the experimental data, as this situation would result in a reduction/disappearance of the dynamic signal with current. Instead, we observe the biggest signals at the highest currents, which most likely is associated with large droplet drift. Then again, the simulations cannot be expected to give a quantitative fit to all data. The model assumes zero temperature and at large currents, this assumption certainly breaks down. Nevertheless, the simulations still reveal intriguing features of the coexisting droplets.

\section*{Conclusions}
We conclude by pointing out that the role of the reference layer has commonly been overlooked in studies of spin-torque stabilized magnetodynamical solitons. Our results show that the RL can not be neglected, particularly not for high currents. The experiments provide compelling evidence of coexisting droplets in the free and reference layer. This double droplet state is also reproduced in micromagnetic simulations. Furthermore, the simulations show that the double droplet stability is tunable by field and frequency, which leads to various correlations between the RL- and FL-droplets. The findings constitute a substantial step towards the comprehensive understanding of magnetic solitons needed to enable practical utilization of the phenomena.

\begin{methods}

\subsection{Device fabrication.} The multilayer stack, consisting of a seed layer, Ta~($4$)/ Cu~($14$) / Ta~($4$) / Pd~($2$), an all-perpendicular pseudo-spin valve structure, [Co~($0.35$) / Pd~($0.7$)]${\times5}$ / Co~($0.35$) / Cu~($5$) / [Co~($0.22$) / Ni~($0.68$)]${\times4}$ / Co~($0.22$), and a cap layer, Cu~($2$) / Pd~($2$), was deposited on a thermally oxidized Si wafer using DC/RF magnetron sputtering (numbers in parentheses are thicknesses in nanometers). In the all-perpendicular pseudo-spin valve structure, the [Co/Pd] multilayer is regarded as the reference layer and the [Co/Ni] stack as the free layer, as presented in Fig.~\ref{fig1}. Using conventional optical lithography and dry etching techniques, 8~$\upmu$m $\times$ 16~$\upmu$m mesas were patterned on the stacked wafer and insulated by a 30-nm-thick SiO$_2$ film using chemical vapor deposition (CVD). Then, electron beam lithography (EBL) and reactive ion etching (RIE) were used to fabricate nanocontacts through the SiO$_2$ on top of each mesa. The nanocontacts had a circular diameter varying from 50 to 150 nm. Finally,  500~nm of Cu followed by 100~nm Au was deposited on top and the contact electrode was produced by lift-off processing. The device used in the measurements displayed in Fig.~\ref{fig1}--\ref{fig4} had a nanocontact with a 60-nm diameter.

\subsection{Magnetic and electrical characterization.} The external fields were swept normal to the thin-film plane. Microwave and dc measurements of the fabricated STNOs were carried out using our custom-built 40-GHz probe station. It allows the manipulation of magnetic field strength, polarity, and direction. The device was connected through a ground--signal--ground (GSG) probe. The direct current, using a Keithley 6221 current source, flowed into the probe (so as the device) through a 40-GHz bias-Tee. The dc voltage was measured with a Keithley 2182 nanovoltmeter. Here we define the negative sign of the applied direct current as the electrons flow from the free to the reference layer. When the current generates enough STT, auto-oscillation arises and emits a microwave signal. This microwave signal was decoupled from the dc voltage via the bias-Tee and then amplified using a low-noise amplifier prior to being recorded by a spectrum analyzer (R$\& $S FSU 20 Hz--67 GHz).

\subsection{Micromagnetic simulations.}
Micromagnetic simulations were performed using the GPU-based open-source MuMax3 code.\cite{Vansteenkiste2014} Default settings were used for the solver, including the time step duration. The STNO was modeled by $512\times512\times3$ cells with a cell size of $3.90625\times3.90625\times3.90625$~nm$^3$. \textit{Region1} was defined as the first $512\times512$ layer and corresponds to the [Co/Pd] reference layer. \textit{Region2} constituted the last layer representing the [Co/Ni] free layer. The middle layer refers to the Cu spacer. The different thicknesses of the FL (RL) layer were accounted for by setting the variable "FreeLayerThickness.SetRegion" to $3.90625$~nm for \textit{Region2} ($7.8125$~nm for \textit{Region1}).

The drive current flow was modeled by a simple cylinder with an 80-nm diameter (NC size) and the Oersted field was calculated and included. Zhang-Li torque was not taken into consideration in the simulations, since the current path was modeled without any xy-component. The interlayer exchange coupling between the RL and FL was set to 0. Magnetic parameters of the FL (RL) were the uniaxial magnetic anisotropy $K_{\mathrm{u}}= 340$~kJ/m$^3$ (375~kJ/m$^3$) as determined by out-of-plane FMR measurements, together with the literature value of the saturation magnetization $M_{\mathrm{s}}=716.2$~kA/m (730~kA/m)\cite{Iacocca2014}. The same standard values were used for both layers: gyromagnetic ratio $\gamma /2 \pi = 28$~GHz/T, exchange stiffness $A_{\mathrm{ex}}=10$~pJ/m, damping constant $\alpha=0.03$, current polarization $P=0.4$, and spin torque asymmetry parameter $\Lambda = 1.3$. To mimic the back-hopping effect, we consider the real-time magnetization of one layer as the polarization layer of the other layer. In other words, the real-time states affect each other through the STT effect and the polarization was updated for each time step. 

Absorbing boundary conditions in the form of a smoothly increasing damping profile were applied to the simulated sample edges to avoid any interference artifacts from spin wave reflection. The applied field and initial magnetization angle were set to 89.7 degrees to mimic uncertainties in the experimental setup and to avoid any singularities associated with using an exact number of 90 degree. The magnetization components shown in Fig.~\ref{fig5}g-j and~\ref{fig6}g-j were calculated over an area of $128\times128$ cells, i.e. $500\times500$~nm$^2$, with the nanocontact in the middle. The NC region thus constitutes about 2\% of the sampled area. The momentary magnetization values were saved every $6$~ps.

\subsection{Data availability.} The data that support the findings of this study are available from the corresponding author upon reasonable request.
\subsection{Code availability.} The code used in this study is available from the corresponding author upon reasonable request.

\end{methods}

\begin{addendum}
 \item[Acknowledgements] Support from the Swedish Research Council (VR), the Swedish Foundation for Strategic Research (SSF), the G\"oran Gustafsson Foundation, and the Knut and Alice Wallenberg Foundation is gratefully acknowledged. S.J. acknowledges the financial support from the Natural Science Foundation of China (Grant 621044196) and Basic Research Programs of Taicang (Grant TC2021JC19) and Chongqing Natural Science Foundation (Grant 2022NSCQ- MSX4891). S.C. acknowledges support from the National Research Foundation of Korea (NRF) grant, funded by the Ministry of Science and ICT, Korea (Grants 2020R1F1- A1049642 and 2022M3F3A2A03014536). The work by OH was funded by the US Department of Energy Office of Science Basic Energy Sciences Division of Materials Science and Engineering. We gratefully acknowledge the computing resources provided on Blues and Bebop, high-performance computing clusters operated by the Laboratory Computing Resource Center at Argonne National Laboratory.
 \item[Author contributions] S.J., S.C., Q.T.L., and H.M. characterized and optimized the material stack. S.J., S.C., Q.T.L., and A.H. and fabricated the devices. S.J and S.C. performed all the microwave and electrical measurements. A.F. and O.H. carried out the micromagnetic simulations. S.C. and J.\AA. initiated and supervised the project. All authors contributed to the data analysis and co-wrote the manuscript.
 \item[Competing Interests] The authors declare that they have no competing interests.
 \item[Correspondence] Correspondence and requests for materials should be addressed to J.\AA{}.~(email: johan.akerman@physics.gu.se), S.C.~(email: sjchung76@knue.ac.kr) or M.A.~(email: martina.ahlberg@physics.gu.se).

\end{addendum}

\section*{References \\}

\end{document}